# Sparse Bayesian Learning for EEG Source Localization

Sajib Saha[1,3], Frank de Hoog[2], Ya.I. Nesterets[1,4], Rajib Rana[5], M. Tahtali[3] and T.E. Gureyev[1,4]

[1] CSIRO Materials Science and Engineering, Clayton, VIC 3168, Australia

[2] CSIRO Computational Informatics, Canberra, ACT

[3] University of New South Wales, Canberra, ACT 2610, Australia

[4] University of New England, Armidale, NSW 2351, Australia

[5] University of Southern Queensland, Australia

## Abstract

**Purpose:** Localizing the sources of electrical activity from electroencephalographic (EEG) data has gained considerable attention over the last few years. In this paper, we propose an innovative source localization method for EEG, based on Sparse Bayesian Learning (SBL).

**Methods:** To better specify the sparsity profile and to ensure efficient source localization, the proposed approach considers grouping of the electrical current dipoles inside human brain. SBL is used to solve the localization problem in addition with imposed constraint that the electric current dipoles associated with the brain activity are isotropic.

**Results:** Numerical experiments are conducted on a realistic head model that is obtained by segmentation of MRI images of the head and includes four major components, namely the scalp, the skull, the cerebrospinal fluid (CSF) and the brain, with appropriate relative conductivity values. The results demonstrate that the isotropy constraint significantly improves the performance of SBL. In a noiseless environment, the proposed method was found to accurately (with accuracy of >75%) locate up to 6 simultaneously active sources, whereas for SBL without the isotropy constraint, the accuracy of finding just 3 simultaneously active sources was <75%.

**Conclusions:** Compared to the state-of-the-art algorithms, the proposed method is potentially more consistent in specifying the sparsity profile of human brain activity and is able to produce better source localization for EEG.

## Introduction:

The problem of source localization in EEG has gained significant attention in recent years because of its potential diagnostic value for epilepsy [1], stroke [2, 3], traumatic brain injury [4] and other brain disorders. Localization of the sources of electrical activity inside the brain is carried out by measuring the scalp potentials produced by the electric activity in the brain (this is modelled using electric current dipoles), and then working back and estimating the dipoles that best fit the measurements. The EEG source localization represents a high-dimensional inverse problem which is severely ill-posed [5] by nature and has an infinite number of solutions [6]. In order to find an appropriate unique solution among the set of possible ones, constraints are introduced into the problem [7]. In the literature, the most commonly used constraint is the minimum-norm constraint [8-10], which finds the solution that best matches the measurements with the smallest $l_2$ residuals. The strength of $l_2$ norm based approaches (including Support Vector Machines [36]) is their low computational cost; however such methods are often criticized for generating very broadly distributed or "smeared" sources in the reconstruction region [11] and for poor performance for multiple simultaneously active sources [12, 13]. Following the discovery by Donoho and Candes et al. [14, 15] that sparsity [1] could enable exact solution of ill-posed problems under certain conditions, there has been a tremendous growth of publications [16-21] on efficient application of sparsity constraints for ill-posed problems. In EEG, many researches have used sparseness constraints in spatial, spatiotemporal and frequency domains [6] to reflect the focal nature of the cortical activity. Techniques relying on sparseness constraint search for a solution vector that not only matches the measurements, but also has as few nonzero entries as possible. Among the list of approaches that provide a sparse solution [22], SBL [23, 24] represents an important family of algorithms. In EEG, where the dictionary matrix is highly coherent, an SBL-based approach, TMSBL (i.e. SBL for multiple measurement vector with temporally correlated source vectors) [25] has been found superior over other state-of-the-art sparsity prior approaches. Despite its success in accurately locating up to 3 active sources for a dictionary matrix of size 80×390 (where 80 is the number of measurements and 390 is the number of unknowns), TMSBL and other similar sparsity prior approaches are still insufficient in most realistic scenarios. For example, it is well known that Brodmann area 17 [26] is related to the human visual activity. For a realistic head model, which contains about 6,000 dipoles [3], with each dipole corresponding to about 5×5×5 mm$^3$ of grey matter, the number of dipoles that belong to that region (Brodmann area 17) is about 54. Relying on the well accepted phenomenon in the EEG literature, that a region of the brain corresponding to a group of dipoles, rather than to a single dipole, tends to be activated during a certain brain activity [5], TMSBL is likely to produce inaccurate results for any group of more than 3 active dipoles.

Assuming typical activation of a group of dipoles, rather than a single dipole for a certain brain activity, and taking into account the severely underdetermined nature of the problem, the sparsity-based reconstruction methodology proposed in this paper groups the dipoles and assumes simultaneous activation of all or majority of the dipoles in the group. In addition, it is assumed that electrical activity inside the human brain is sparse when represented in terms of activity of groups of dipoles. While, grouping based on functionality would be the most appropriate, the full functionality of different parts of the human brain is still not known completely and the grouping of dipoles in the present work is anatomical [27] based on AAL (Automated Anatomical Labelling) scheme rather than on a functional classification of the human cerebral cortex. It is worth mentioning that the AAL map [27] is frequently used in fMRI to describe out the region of interest [28], which indicates the potential suitability of such segmentation of the brain as a tentative basis for functional classification relevant to EEG. In this paper we consider each group of dipoles to have the same magnitude and orientation of their dipole moments.

In addition to grouping dipoles, in this paper we propose an enhanced version of the Sparse Bayesian Learning methodology for EEG source localization, that takes into account the fact that when a brain region is active there is likely to be a contribution to the scalp electric potentials from each of the components of the current density vector. Existing works based on Sparse Bayesian Learning in EEG do not make a distinction between the treatment of spatial regions and the directional components of the current distribution within those regions. Specifically, each of the directional components of the current are assumed to be sparse independently. However, it is more logical to assume that when a brain region is active there is contribution to the potential from each of the components of the current density vector. In the absence of prior information about the current distribution, a better choice is an uninformative prior that the activity in each region is isotropic and this is what is done in the present paper. An advantage of this approach is that it requires the estimation of one third of the number of hyper-priors compared with putting an independent prior in each of the three spatial directions. In our simulations, each group of dipoles is assumed to have the same magnitude. While the effect of the partial activation of the groups is reported, the effect of different orientation of dipoles in a group has not been investigated yet.

---

[1] A signal $x$ is called $S$-sparse if it has only $S$ nonzero elements. However, natural signals are rather compressible. A signal $x$ is designated compressible if it has only a small proportion of large coefficients when the signal is transformed into a suitable domain such as, Haar, Fourier etc. Mathematically, a signal is compressible if the coefficients decay obeys the power law[15].

**Mathematical Formulation of EEG:**

The model used for the forward and inverse imaging problems in EEG has the following form:

$$\boldsymbol{\Phi} = \boldsymbol{KJ} + \boldsymbol{n}. \tag{1}$$

Here $\boldsymbol{\Phi} \in \mathbb{R}^{N_E \times 1}$ is a vector of the scalp electric potentials measured by the $N_E$ electrodes with respect to a reference electrode, $\boldsymbol{J} \in \mathbb{R}^{3N_V \times 1}$ is the primary or impressed current dipole vector, where $N_V$ is the number of dipole locations in the brain, with each dipole current vector having three independent components corresponding to the usual Cartesian coordinates in 3D space, $\boldsymbol{K} \in \mathbb{R}^{N_E \times 3N_V}$ is the so-called lead field matrix and $\boldsymbol{n}$ is the noise vector. The lead field matrix has the following structure.

$$\boldsymbol{K} = \begin{bmatrix} K_{1,x}^1 & K_{1,y}^1 & K_{1,z}^1 & ... & K_{N_V,x}^1 & K_{N_V,y}^1 & K_{N_V,z}^1 \\ K_{1,x}^2 & K_{1,y}^2 & K_{1,z}^2 & ... & K_{N_V,x}^2 & K_{N_V,y}^2 & K_{N_V,z}^2 \\ & & & ... & & & \\ & & & ... & & & \\ K_{1,x}^{N_E} & K_{1,y}^{N_E} & K_{1,z}^{N_E} & ... & K_{N_V,x}^{N_E} & K_{N_V,y}^{N_E} & K_{N_V,z}^{N_E} \end{bmatrix}$$

where $K_{p,D}^l$ is the scalp electric potential at the $l$th electrode, due to a unit strength dipole with orientation $D \in (x, y, z)$ that is located in the $p$th voxel.

**Sparse Bayesian Learning:**

Sparse Bayesian Learning was initially proposed for regression and classification by Tipping [24] in the machine learning context. In [22], Wipf et al. applied SBL for the sparse signal recovery problem. The idea of SBL is to find $\boldsymbol{J}$ through Maximum a Posteriori (MAP) estimate [23, 24]. In common with other methods for sparse signal recovery [29-31], SBL assumes that the components of the noise vector are independent, identically distributed Gaussian random variables and this leads to the following probability density function for $\boldsymbol{\Phi}$ [22]

$$p(\boldsymbol{\Phi}|\boldsymbol{J}, \sigma^2) = (2\pi\sigma^2)^{-\frac{N_E}{2}} \exp(-\frac{1}{2\sigma^2}||\boldsymbol{\Phi} - \boldsymbol{KJ}||^2).$$

Obtaining maximum likelihood estimates for $\boldsymbol{J}$ is equivalent to finding the minimum-norm solution to (1) and such solutions are well known to produce non-sparse representations [22]. To alleviate this problem, SBL incorporates a prior distribution for the components of the current density that encourages sparsity. Typically, this is achieved by assuming that that the components are independent mean zero random variables, each with a different variance. That is,

$$p(\boldsymbol{J}, \boldsymbol{\omega}) = \prod_{h=1}^{3N_V} (2\pi\omega_h)^{-\frac{1}{2}} \exp(-\frac{J_h{}^2}{2\omega_h}).$$

However, in the absence of prior information about the current distribution, a better choice is an uninformative prior where the activity in each region is isotropic. This is achieved by assuming that, for a given region, the variances of the components is the same. That is, $\boldsymbol{\omega} = (\gamma_1, ..., \gamma_{N_V})^T \otimes [1 \ 1 \ 1]^T$ , where $\boldsymbol{\gamma} = (\gamma_1, ..., \gamma_{N_V})$ , is a vector of nonnegative hyperparameters and $\otimes$ is the Kronecker product operator. This approach allows one to achieve a three-fold reduction of the number of hyperparameters.

The hyperparameters (along with the error variance $\sigma^2$) are estimated from the data by marginalizing over the weights and then performing maximum likelihood (ML) optimization. Following Zhang et al. [23] the marginalized probability density function (pdf) is given by

$$p(\boldsymbol{\Phi}, \boldsymbol{\omega}, \sigma^2) = \int p(\boldsymbol{\Phi}|\boldsymbol{J}, \sigma^2) \, p(\boldsymbol{J}, \boldsymbol{\omega}) d\boldsymbol{J}$$

$$= (2\pi)^{-\frac{N_E}{2}} |\boldsymbol{\Psi_\Phi}|^{-\frac{1}{2}} \exp[-\frac{1}{2}\boldsymbol{\Phi}^T \boldsymbol{\Psi_\Phi}^{-1} \boldsymbol{\Phi}] \quad (2)$$

where $\boldsymbol{\Psi_\Phi} \triangleq (\sigma^2 \boldsymbol{I} + \boldsymbol{K}\boldsymbol{\Psi_0}\boldsymbol{K}^T)$, and

$$\boldsymbol{\Psi}_0 = \begin{bmatrix} \gamma_1 & & \\ & \ddots & \\ & & \gamma_{N_V} \end{bmatrix} \otimes \begin{bmatrix} 1 & & \\ & 1 & \\ & & 1 \end{bmatrix}. \quad (3)$$

Using the Bayes rule we obtain the posterior density of $\boldsymbol{J}$ which is also Gaussian,

$$p(\boldsymbol{J}|\boldsymbol{\Phi}, \boldsymbol{\gamma}, \sigma^2) = \mathcal{N}_J(\boldsymbol{\mu}_J, \boldsymbol{\Psi}_J),$$

with mean

$$\boldsymbol{\mu}_J = \frac{1}{\sigma^2} \boldsymbol{\Psi}_J \boldsymbol{K}^T \boldsymbol{\Phi} \quad (4)$$

and covariance matrix

$$\boldsymbol{\Psi}_J = (\boldsymbol{\Psi}_0{}^{-1} + \frac{1}{\sigma^2} \boldsymbol{K}^T \boldsymbol{K})^{-1}$$

$$= \boldsymbol{\Psi}_0 - \boldsymbol{\Psi}_0 \boldsymbol{K}^T (\sigma^2 \boldsymbol{I} + \boldsymbol{K}\boldsymbol{\Psi}_0\boldsymbol{K}^T)^{-1} \boldsymbol{K}\boldsymbol{\Psi}_0. \quad (5)$$

So given the hyperparameters, (i.e. $\boldsymbol{\gamma}$ and $\sigma^2$), the MAP (maximum a posteriori) estimate of $\boldsymbol{J}$ is given by

$$\hat{\boldsymbol{J}} \triangleq \boldsymbol{\mu}_J = (\sigma^2 \boldsymbol{\Psi}_0{}^{-1} + \boldsymbol{K}^T \boldsymbol{K})^{-1} \boldsymbol{K}^T \boldsymbol{\Phi}$$

$$= \boldsymbol{\Psi}_0 \boldsymbol{K}^T (\sigma^2 \boldsymbol{I} + \boldsymbol{K}\boldsymbol{\Psi}_0\boldsymbol{K}^T)^{-1} \boldsymbol{\Phi}.$$

Following Zhang [23], to find the hyperparameters $\Theta = \{\boldsymbol{\gamma}, \sigma^2\}$, we employ the Expectation Maximization (EM) method to maximize $p(\boldsymbol{\Phi}, \Theta)$.

The EM formulation proceeds by treating $\boldsymbol{J}$ as hidden variables and then maximizing

$$Q(\Theta) = E_{\boldsymbol{J}|\boldsymbol{\Phi},\Theta^{(old)}}[\log p(\boldsymbol{\Phi}, \boldsymbol{J}, \Theta)]$$

$$= E_{\boldsymbol{J}|\boldsymbol{\Phi},\Theta^{(old)}}[\log p(\boldsymbol{\Phi}|\boldsymbol{J}, \sigma^2)] + E_{\boldsymbol{J}|\boldsymbol{\Phi},\Theta^{(old)}}[\log p(\boldsymbol{J}, \gamma_1, \dots, \gamma_{N_V})] \tag{6}$$

where $\Theta^{(old)}$ denotes the estimated hyperparameters in the previous iteration.

To estimate $\boldsymbol{\gamma}$, we notice that the first term in (6) is unrelated to $\boldsymbol{\gamma}$ and thus the $Q$ function can be simplified to

$$Q(\boldsymbol{\gamma}) = E_{\boldsymbol{J}|\boldsymbol{\Phi},\Theta^{(old)}}[\log p(\boldsymbol{J}, \boldsymbol{\gamma})].$$

From the analysis in [23], it follows that

$$\log p(\boldsymbol{J}, \boldsymbol{\gamma}) \propto -\frac{1}{2}\log(|\boldsymbol{\Psi}_0|) - \frac{1}{2}\boldsymbol{J}^T(\boldsymbol{\Psi}_0^{-1})\boldsymbol{J},$$

which results in

$$Q(\boldsymbol{\gamma}) \propto -\frac{1}{2}\log(|\boldsymbol{\Psi}_0|) - \frac{1}{2}\text{Tr}[\boldsymbol{\Psi}_0^{-1}(\boldsymbol{\Psi}_J + \boldsymbol{\mu}_J\boldsymbol{\mu}_J^T)], \tag{7}$$

where $\boldsymbol{\mu}_J$ and $\boldsymbol{\Psi}_J$ are evaluated using equations (4) and (5) respectively, and the old estimated hyperparameter $\Theta^{(old)}$.

From equation (3) we have

$$\boldsymbol{\Psi}_0 = \text{diag}(\boldsymbol{\gamma}) \otimes \begin{bmatrix} 1 & & \\ & 1 & \\ & & 1 \end{bmatrix} \text{ yielding } \frac{\partial \boldsymbol{\Psi}_0}{\partial \gamma_i} = \text{diag}(\boldsymbol{e}) \otimes \begin{bmatrix} 1 & & \\ & 1 & \\ & & 1 \end{bmatrix}, \text{ where } \boldsymbol{e}_i \in \mathbb{R}^{N_V} \text{ is}$$

the unit vector with 1 in the $i$th component.

Now the derivative of (7) with respect to $\gamma_i (\forall_i \in 1\!:\!N_V)$ is given by

$$\frac{\partial Q}{\partial \gamma_i} = -\frac{3}{2\gamma_i} + \frac{1}{2\gamma_i^2}\text{Tr}[(\text{diag}(\boldsymbol{e}) \otimes \begin{bmatrix} 1 & & \\ & 1 & \\ & & 1 \end{bmatrix})(\boldsymbol{\Psi}_J + \boldsymbol{\mu}_J(\boldsymbol{\mu}_J)^T)(\text{diag}(\boldsymbol{e}) \otimes \begin{bmatrix} 1 & & \\ & 1 & \\ & & 1 \end{bmatrix})]$$

So the learning rule for $\gamma_i$ becomes

$$\gamma_i \leftarrow \frac{1}{3}\text{Tr}[(\text{diag}(\boldsymbol{e}) \otimes \begin{bmatrix} 1 & & \\ & 1 & \\ & & 1 \end{bmatrix})(\boldsymbol{\Psi}_J + \boldsymbol{\mu}_J(\boldsymbol{\mu}_J)^T)(\text{diag}(\boldsymbol{e}) \otimes \begin{bmatrix} 1 & & \\ & 1 & \\ & & 1 \end{bmatrix})].$$

To learn $\sigma^2$, the $Q$ function is simplified to

$$Q(\sigma^2) = E_{\boldsymbol{J}|\boldsymbol{\Phi};\Theta^{(old)}}[\log p(\boldsymbol{\Phi}|\boldsymbol{J}; \sigma^2)]$$

$$\propto -\frac{N_E}{2}\log(\sigma^2) - \frac{1}{2\sigma^2}E_{\boldsymbol{J}|\boldsymbol{\Phi};\Theta^{(old)}}[||\boldsymbol{\Phi} - \boldsymbol{K}\boldsymbol{J}||_2^2]$$

$$= -\frac{N_E}{2}\log(\sigma^2) - \frac{1}{2\sigma^2}\left[\left|\left|\boldsymbol{\Phi} - \boldsymbol{K}\boldsymbol{\mu_J}\right|\right|_2^2 + \mathrm{Tr}(\boldsymbol{\Psi_J}\boldsymbol{K^T}\boldsymbol{K})\right]$$

$$= -\frac{N_E}{2}\log(\sigma^2) - \frac{1}{2\sigma^2}\left[\left|\left|\boldsymbol{\Phi} - \boldsymbol{K}\boldsymbol{\mu_J}\right|\right|_2^2 + \sigma^2[N_V - \mathrm{Tr}(\boldsymbol{\Psi_J}\boldsymbol{\Psi_0}^{-1})]. \quad (8)$$

By taking the derivatives of equation (8) over $\sigma^2$ and setting it to zero the $\sigma^2$ learning rule becomes

$$\sigma^2 \leftarrow \frac{\left|\left|\boldsymbol{\Phi} - \boldsymbol{K}\boldsymbol{\mu_J}\right|\right|_2^2 + \sigma^{(old)^2}[N_V - \mathrm{Tr}(\boldsymbol{\Psi_J}\boldsymbol{\Psi_0}^{-1})]}{N_E}$$

where $\sigma^{(old)^2}$ denotes the estimated $\sigma^2$ on the previous iteration.

**Data Model and Assumptions:**

A realistic head model was obtained from segmentation of MRI images of the head and includes four major compartments, namely scalp, skull, cerebrospinal fluid (CSF) and brain, with the following relative conductivity values [5]: $\sigma_{scalp}=1$, $\sigma_{skull}=0.05$, $\sigma_{CSF}=5$, $\sigma_{brain}=1$. The source space was constructed by dividing the head model into 5×5×5 mm³ cubes and considering possible current dipoles only at the centre of those cubes that consisted of at least 60% of gray matter. This segmentation procedure resulted in 6203 dipole positions. In order to implement the proposed sparsity criteria, the 6203 dipoles were grouped based on anatomical structure of the cerebral cortex. Since Automated Anatomical Labeling (AAL) [27] is frequently used in functional MRI (fMRI) to locate functional activity inside human brain, we used the AAL template from the MRIcro [32] package. The template consists of 116 areas in the standard MNI space. Thus while forming clusters each of the 6203 dipoles was assigned a tag based on the closely located anatomical area. Dipoles with the same tag formed a cluster. Theoretically, we expected to form 116 clusters. However in our case we were able to form only 113 clusters due to the coarse sampling of the head model in EEG compared to fMRI and also due to the percentage of gray matter associated with a dipole; which resulted in zero dipoles in small anatomical areas. Assuming uniform activity across a whole area/group, the resultant 113 clusters which we called Functional Zones were then used to recompute the previously calculated lead field matrix from $\boldsymbol{K} \in \mathbb{R}^{N_E \times (18609)}$ to $\boldsymbol{K_{FZ}} \in \mathbb{R}^{N_E \times 339}$, by taking the average of all the lead field values belonging to each group.

**Localizing the Sources of Electrical Activity:**

To localize the sources of electrical activity we solve the inverse problem based on the proposed Sparse Bayesian Learning and with 113 Functional Zones. The solution $\hat{\boldsymbol{J}} \in \mathbb{R}^{339 \times 1}$ is a -component vector representing current sources at 113 locations within the brain volume with three directional (i.e. $x$, $y$, $z$ directions) components per location. The $x$, $y$ and $z$ components are used to calculate the magnitude of the current density for each of the Functional Zones. From the magnitudes of the current density of the Functional Zones, the maximum magnitude, $|\boldsymbol{J}|_{\mathbf{max}}$ is determined. Any Functional Zone with a magnitude larger than or equal to $t|\boldsymbol{J}|_{\mathbf{max}}$, is considered to be active. The threshold, $t$ is experimentally set to 1/3 of $|\boldsymbol{J}|_{\mathbf{max}}$.

**Experimental Analysis:**

Experiments were conducted by varying the number of simultaneously active Functional Zones for the EEG headset configuration shown below.

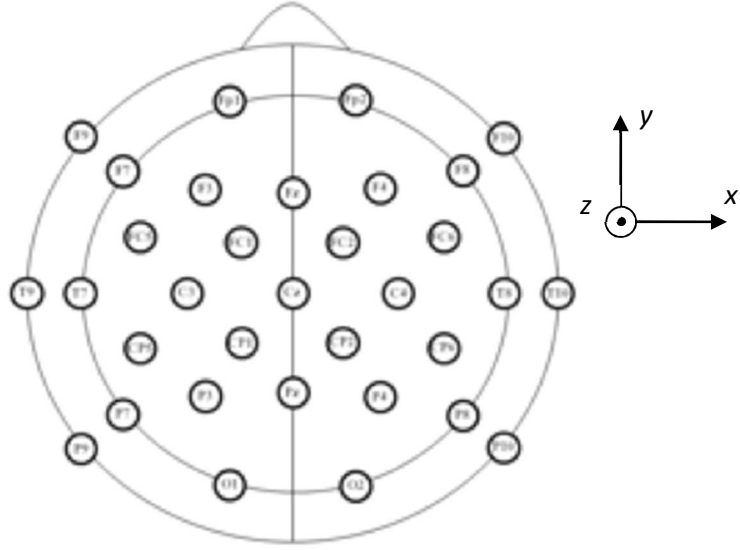

Figure 1: Schematic representation of the electrodes positions in the 33-electrodes setup.

We analyzed the performance of the proposed method against TMSBL [23] method, where in both cases we grouped the dipoles into Functional Zones, to reduce the dimensionality of the problem and to better specify the sparsity profile.

Error Distance (ED) [33] was used to analyse the reconstruction performance. The error distance between the actual and the estimated source locations is defined as

$$ED = \frac{1}{N_I} \times \sum_{i \in I}^{N_I} min_l \left|\left|\boldsymbol{r}_i - \boldsymbol{s}_l\right|\right| + \frac{1}{N_L} \times \sum_{l \in L}^{N_L} min_i \left|\left|\boldsymbol{s}_l - \boldsymbol{r}_i\right|\right| \qquad (7)$$

Here $\boldsymbol{s}_l$ and $\boldsymbol{r}_i$ are the actual and estimated source locations respectively. $N_I$ and $N_L$ are the total numbers of estimated and the undetected sources respectively. The first term of equation (7) calculates the mean of the distance from each estimated source to its closest real source, and the corresponding real source is then marked as detected. All the undetected real sources made up the elements of the data set $L$ and thus the second term of the equation calculates the mean of the distance from each of the undetected sources to the closest estimated source.

**A) Noise-Free Simulations:**

**i) Localization Error - Single Source Activation:**

First, a single Functional Zone was considered active and the corresponding potentials on the electrodes were calculated based on the averaged lead field matrix $\boldsymbol{K_{FZ}}$. The experiment was conducted for all Functional Zones activated sequentially one at a time. Each activated Functional Zone had the same magnitude but different orientations (chosen randomly) of the dipole moment (i.e. the orientation of the average dipole moment). For this experiment, each Functional Zone was represented by its centroid $\boldsymbol{s}_i (\forall_i \in 1 : \boldsymbol{FZ})$. For an active source position $\boldsymbol{s}_i$ we claimed a "success" if the error distance was zero. The success rate was computed as $\frac{\sum_{i=1}^{AZ}(ED^i==0)}{FZ}$. We computed mean error distance, $E = \sum_{i=1}^{FZ} ED^i / N_{unsuccessful}$ for unsuccessful cases as localization error. The experiment was repeated 100 times. For each run we computed the success rate, the mean error distance (considering only unsuccessful cases) and the standard deviation. Table 1 shows the average of 100 such findings.

Table 1: Success rate and the localization error analysis for unsuccessful cases for one Active Zone (67 Functional Zones in total).

|  | Proposed method | TMSBL |
|---|---|---|
| Success rate | 1.00 | 0.96 |
| Mean error distance (for unsuccessful cases) (mm) | 0 | 20.08 |
| Standard deviation of the mean error distance (for unsuccessful cases) (mm) | 0 | 10.71 |

## ii) Localization Error – Multiple Sources Activation:

In this case $S$ ($S > 1$) Functional Zones were activated simultaneously. From the total of $C_S^{113} = \frac{113!}{S!(113-S)!}$ possible combinations of Functional Zones one combination was chosen randomly. Then we computed the error distance between the actual and reconstructed signal and claimed a "success" if the computed error distance was zero. We did the experiment 1000 times (varying the combinations of activated Functional Zones and the orientation of the average dipole moment) and the success rate was computed over these 1000 runs. For the unsuccessful cases we computed the mean error distance, $E = \frac{\sum_{i=1}^{1000} ED^i}{N_{unsuccessful}}$ and the standard deviation of the mean error distance. We did the same for each values of $S$. Figure 2 shows the success rates as a function of $S$. Table 2 summarizes the localization errors for unsuccessful cases.

As expected, both for the proposed method and for TMSBL the success rate decreases with the increased number of simultaneously active zones. In more than 75% of the cases, the proposed method obtained accurate localization for up to 6 simultaneously active Functional Zones. This degree of accuracy not achieved by TMSBL, even for the much simpler problem of locating just 3 simultaneously active Functional Zones.

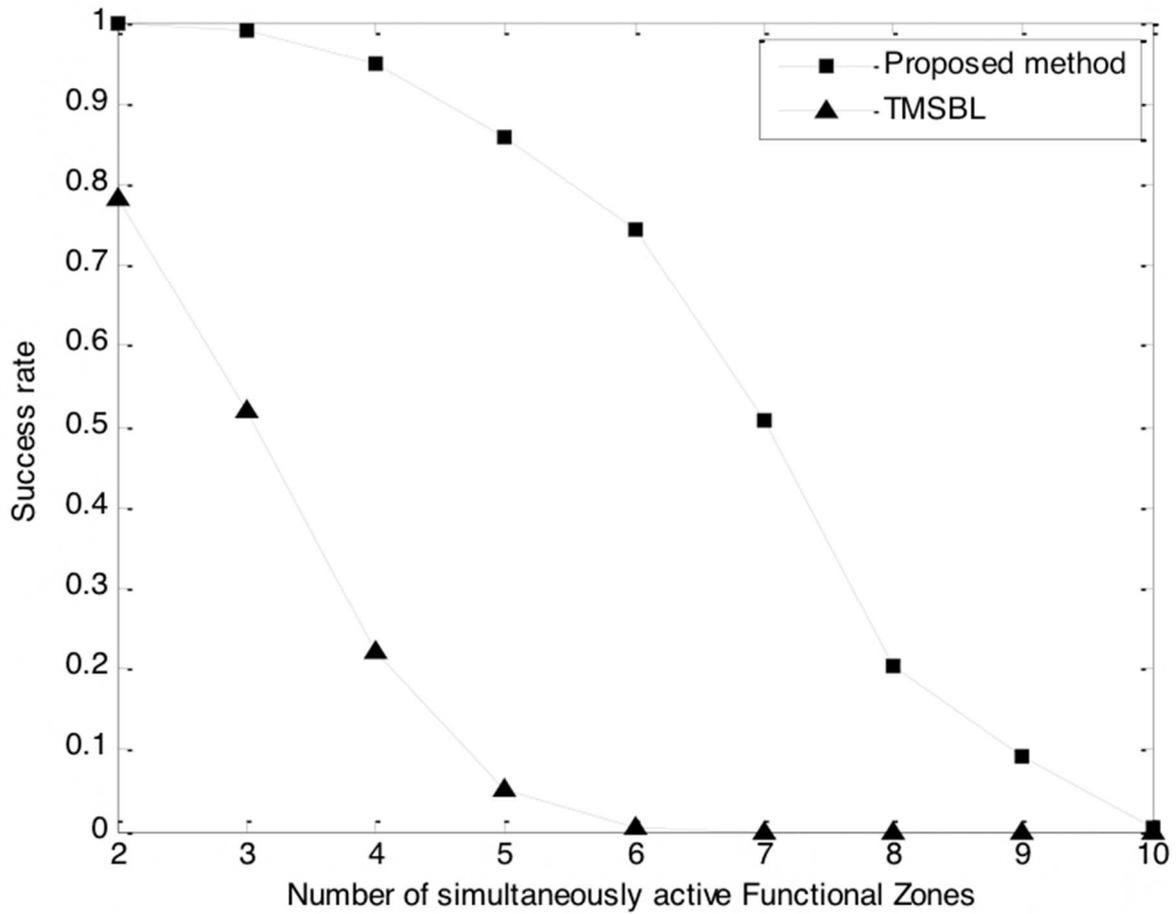

Figure 2: Success rate of the reconstruction in regard to number of simultaneously active areas. Here "success" meant that all the activated Functional Zones were exactly located in the reconstructed signal.

Table 2: Localization error analysis for unsuccessful cases, where "success" meant that all the activated Functional Zones were exactly located in the reconstructed signal.

| Number of simultaneously active areas | Proposed method | | TMSBL | |
|---|---|---|---|---|
| | Mean error distance (mm) | Standard deviation (mm) | Mean error distance (mm) | Standard deviation (mm) |
| 2 | 18.2119 | 15.2933 | 23.2785 | 19.3761 |
| 3 | 28.1601 | 23.5682 | 25.0346 | 21.4945 |
| 4 | 27.7569 | 21.8031 | 31.9021 | 21.7739 |
| 5 | 39.4475 | 24.3093 | 41.5874 | 22.0897 |
| 6 | 41.2074 | 23.8359 | 49.3348 | 21.6766 |
| 7 | 44.1597 | 21.5332 | 54.3522 | 19.9626 |
| 8 | 46.3797 | 18.8601 | 57.5860 | 17.7100 |
| 9 | 47.9485 | 17.2700 | 58.9300 | 17.2988 |
| 10 | 51.2658 | 16.6363 | 59.1794 | 16.4202 |

### iii) Partially activated Functional Zone:

This experiment was designed to analyse the performance of the localization method when only a fraction of the Functional Zone rather than the whole Functional Zone was active. For each of the activated Functional Zone a specified percentage of the dipoles belonging to that Functional Zone was activated. One dipole of the considered Functional Zone was chosen randomly and the rest of the active dipoles (based on the specified percentage of activated dipoles) were adjacent to the chosen dipole. The complete Lead field matrix $K$ was used for the forward problem (i.e. to generate $\Phi$), whereas $K_{FZ}$ was used for the inverse problem. While Functional Zone(s) can have random direction of activity, all the dipoles within a given Functional Zone were considered to have the same orientation. The experiment was conducted for all the Functional Zones ($1:FZ$) activated sequentially, one at a time. In this case we claimed a "success" if the activated Functional Zone was exactly detected in the reconstructed signal (i.e. if the error distance was zero). The success rate was computed as $\frac{\sum_{i=1}^{FZ}(ED^i==0)}{FZ}$ and mean error distance, $E$ was computed as $\sum_{i=1}^{FZ} ED^i / N_{unsuccessful}$ considering only unsuccessful cases. The whole experiment was conducted 10 times and for each run we computed the success rate and the mean error distance (considering only unsuccessful cases). The results shown in Table 3 are the average over 10 such findings.

Table 3: Success rate as a function of the activated percentage of a Functional Zone.

| Percentage of the active area | Success rate | | Mean error distance (for unsuccessful cases) (mm) | |
|---|---|---|---|---|
| | Proposed method | TMSBL | Proposed method | TMSBL |
| 100 % | 1 | 0.9668 | 0 | 19.0342 |
| 90 % | 0.9912 | 0.9313 | 10.9501 | 18.2507 |
| 80 % | 0.9209 | 0.8063 | 13.8112 | 15.6526 |
| 70 % | 0.8419 | 0.6972 | 15.2404 | 16.2270 |
| 60 % | 0.7022 | 0.5708 | 14.8319 | 16.4311 |
| 50 % | 0.6324 | 0.4902 | 15.3301 | 16.2450 |

From the results it is clear that when the whole area of the considered zone is active it is very likely that it will be exactly localized in the reconstructed signal. As soon as the percentage of the active area decreases the chances of inexact localization increase. Another important observation is that when the percentage of activation decreases one could have more than one active Functional Zone detected in the reconstructed signal, however it is very likely that the activity maxima will coincide with the actual activated Functional Zone. In order to verify this claim, rather than computing the error distance as in Table 3, we considered only one Functional Zone having the maximum magnitude in the reconstructed signal and then computed the Euclidian distance between the actual and estimated Functional Zone. The findings are shown in Table 4.

Table 4: Success rate for percentage of the Functional Zone. The results shown here is the average over 10 runs.

| Percentage of the active area | Success rate | | Mean Euclidian distance (unsuccessful cases) (mm) | |
|---|---|---|---|---|
| | Proposed method | TMSBL | Proposed method | TMSBL |
| 100 % | 1 | 0.9735 | 0 | 18.9541 |
| 90 % | 0.9999 | 0.9636 | 7.1505 | 17.2557 |
| 80 % | 0.9801 | 0.9383 | 16.8112 | 17.6526 |
| 70 % | 0.9451 | 0.8706 | 18.2404 | 17.2270 |
| 60 % | 0.8601 | 0.7996 | 19.8319 | 19.4311 |
| 50 % | 0.8210 | 0.7352 | 20.3301 | 21.2450 |

Since clustering based on AAL produces non-uniformly sized clusters, for the large clusters ( having >100 dipoles), rather than specifying the active area as percentage of whole area, we specify the active area in terms of the number of dipoles. One dipole of the considered Functional Zone was chosen randomly and the rest of the active dipoles were adjacent to the chosen dipole. We considered only one Functional Zone having the maximum magnitude in the reconstructed signal and then computed the Euclidian distance between the actual and estimated Functional Zone. The findings are shown in Table 5.

Table 5: Success rate for partially activated (defined in terms of the number of active dipoles) Functional Zone. The results shown here is the average over 10 runs.

| Number of active dipoles | Success rate | | Mean Euclidian distance (unsuccessful cases) (mm) | |
|---|---|---|---|---|
| | Proposed method | TMSBL | Proposed method | TMSBL |
| 10 | 0.4047 | 0.3501 | 30.0219 | 31.2713 |
| 20 | 0.4952 | 0.4239 | 26.3471 | 27.9808 |
| 30 | 0.5433 | 0.4555 | 25.3709 | 28.0677 |

**B) Experiment with Noisy Data:**

All of the preceding experiments were performed in the absence of noise in the simulated EEG data. Experimental data will inevitably be contaminated by "noise" from various sources, including measurement noise and background brain activity [34]. Here we investigated the effect of simulated pseudo-random measurement noise superimposed on the measured scalp potentials.

**i) Localization Error - Single Source Activation:**

First, a single Functional Zone was considered active and the corresponding potentials on the electrodes were calculated based on $K_{FZ}$. We then added variable amounts of noise to each of the electrode potentials. The noise vector $\boldsymbol{n}$ was generated based on the SNR (Signal to Noise Ratio) defined below.

$$\text{SNR (in dB)} = 20 \log_{10} \frac{||KJ||_2}{<||n||_2>} \tag{8}$$

where $<\ldots>$ designates the mean value over a statistical ensemble.

In this study we increased the SNR from 5 dB to 30 dB with 5dB increments. The experiment was conducted for all Functional Zones $s_i (\forall_i \in 1:FZ)$ activated sequentially one at a time and for different levels of SNR. Localization of the Functional Zone was performed according to the algorithm specified above in section 'Localizing the Sources of Electrical Activity'. We computed the success rate (computed as $\frac{\sum_{i=1}^{AZ}(ED^i==0)}{FZ}$), the mean error distance (considering only unsuccessful cases). Since we were adding random (normally distributed) noise, we did 100 repetitions of the experiment. Table 6 shows the average of 100 such findings.

Table 6: Success rate and localization error (i.e. error distance) analysis with different levels of SNR, where "success" meant that activated Functional Zone was exactly located in the reconstructed signal

| SNR level | Success rate | | Mean localization error (unsuccessful cases) (in mm) | |
|---|---|---|---|---|
| (in dB) | Proposed method | TMSBL | Proposed method | TMSBL |
| 30 | 0.5981 | 0.3651 | 28.1613 | 31.6501 |
| 25 | 0.5191 | 0.2051 | 28.8527 | 33.3575 |
| 20 | 0.4539 | 0.1763 | 31.1621 | 40.9204 |
| 15 | 0.2610 | 0.1070 | 38.1996 | 44.8093 |
| 10 | 0.1097 | 0.0104 | 53.8633 | 55.7988 |
| 5 | 0.0199 | 0.0095 | 64.0945 | 66.9675 |

**ii) Localization Error – Multiple Source Activation:**

In this case $S$ ($S > 1$) Functional Zones were activated simultaneously. From the total of $C_S^{113}$ possible combinations of Functional Zones, one combination was chosen randomly. Then we computed the error distance between the actual and reconstructed signal as localization error. We performed the experiment 1000 times (varying the combinations of activated Functional Zones and the orientation of the average dipole moment) and the mean error distance was computed over these 1000 runs. We did the same for each values of S. For each value of $S$ the experiment was performed in the presence of 20 dB noise as specified by equation (8). Since we were adding random amount of noise, we did 100 repetitions of the experiment. Figure 3 shows the average of 100 such findings.

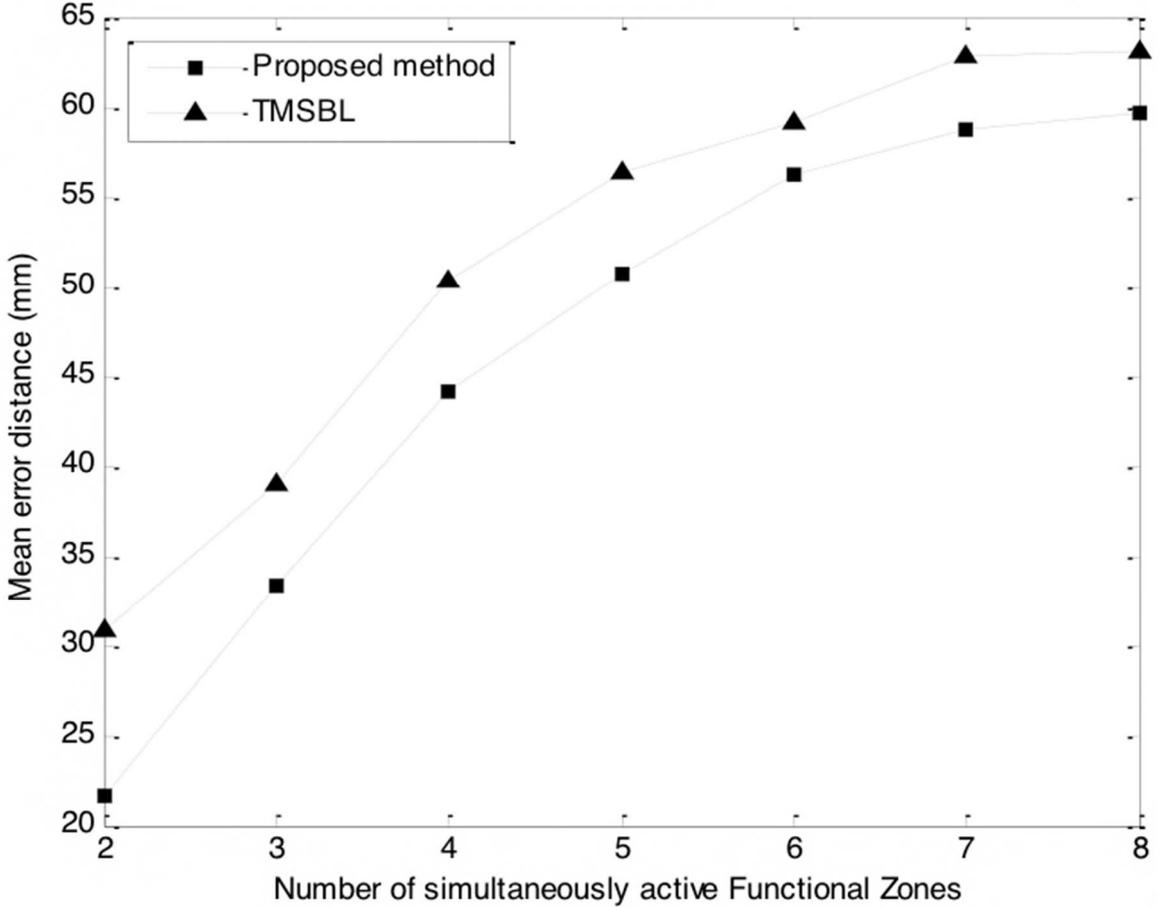

Figure 3: Localization error analysis for simultaneously activated Functional Zones in the presence of 20 dB noise.

## C) Experiment – Simulation Follows the Assumption:

The experiment was conducted for $S$ ($S \geq 1$) simultaneously active Functional Zones, considering two different cases and without the presence of noise. In one case (case I), while generating the forward problem we exactly followed the assumption, i.e. for each activated Functional Zone all three components of the average dipole moment were sampled from Gaussian distribution having zero mean and constant variance. In the other case (case II), we did not necessarily follow the assumption, i.e. for each activated Functional Zone all the three components of the average dipole moment were chosen randomly. For both of the cases the proposed method was used to solve the inverse problem. For a stated $S$, from the total of $C_S^{113} = \frac{113!}{S!(113-S)!}$ possible combinations of Functional Zones one combination was chosen randomly. Then we computed the error distance between the actual and reconstructed signal and claimed a "success" if the computed error distance was zero. We did the experiment 1000 times (varying the combinations of activated Functional Zones and the orientation of the average dipole moment) and the success rate was computed over these 1000 runs. Figure 4 shows the findings.

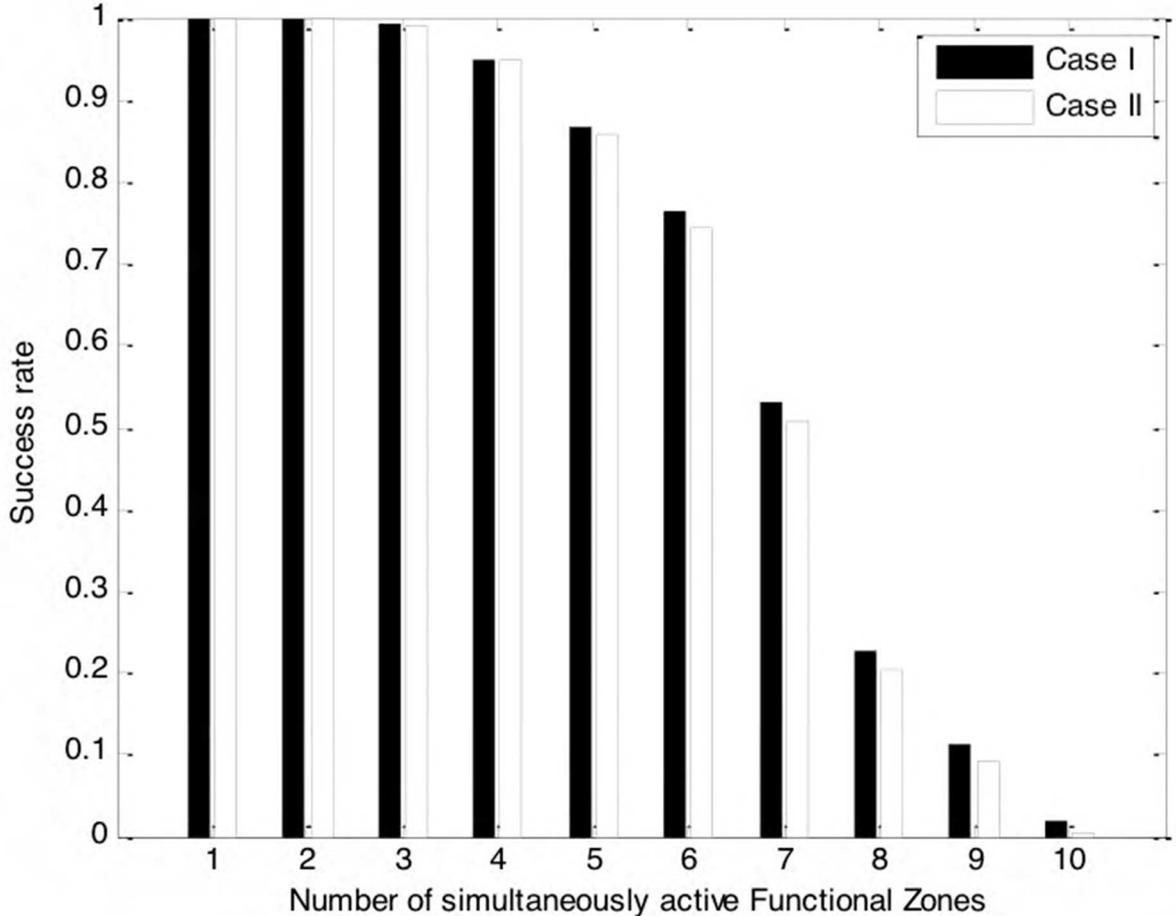

Figure 4: Success rate of the reconstruction as a function of the number of simultaneously active areas. Here "success" meant that all the activated Functional Zones were exactly located in the reconstructed signal. The data are based on the average of 1000 runs for each number of simultaneously active Functional Zone.

**Conclusions:**

By exploiting the well accepted phenomenon that usually a group of adjacent dipoles rather than a single dipole becomes active [35], the proposed approach groups the dipoles inside the human brain on the anatomical/functional basis to reduce the severe underdetermined nature of the EEG problem and thereby achieve better source localization. The grouping of dipoles into several Functional Zones not only reduces the dimensionality of the problem, but also ensures potentially more realistic specification of the sparsity profile. Following the demonstration by Zhang [25] that SBL algorithms are superior to other state-of-the-art methods in EEG, where the dictionary matrix (i.e. lead field matrix) is highly coherent, we have used SBL for solving the localization problem. We also proposed an enhanced version of the SBL algorithm based on an uninformative isotropic prior distribution for each current dipole. While the proposed version of the SBL algorithm this isotropy to solve the inverse problem, we did not necessarily follow this assumption in the simulation to generate the forward data for the experiment in section -A and section-B of the 'Experimental Analysis'. As can be seen from the section-C of the 'Experimental Analysis' that when the forward problem exactly follows the assumption, one might expect even more improvement in the results.

For comparison, the TMSBL [25] code implemented in MATLAB was downloaded from the author's website. Though TMSBL code had the option to incorporate temporal measurements also, in this work we did not consider any temporal measurement neither for TMSBL nor for the proposed method. Out of the two noise variance learning rules that are available in the MATLAB package of TMSBL, we used the one that follows the derivation of [23], both for TMSBL and for the proposed case.

We evaluated the proposed method by varying the number of simultaneously activated Functional Zones and in regard to different levels of noise, using a realistic head model. The results indicate that the proposed method ensures better source localization. In a noiseless environment, the proposed method was found to accurately (with accuracy of >75%) locate up to 6 simultaneously active Functional Zones, whereas for TMSBL, even for 3 simultaneously active Functional Zones, the accuracy was <75%. Also in the presence of noise, the proposed method was found to be more robust against the TMSBL method.

Since the relevant details of the human brain activity are still not known precisely, it is particularly important to know what percentage of the Functional Zone needs to be activated for successful localization of that Zone. Experiments reveal that even with 50% activation, of the considered Functional Zone, there is 80% chance that the activated Functional Zone will be accurately localized by the proposed method.

One fundamental question remains; how meaningful the AAL-based grouping is for the proposed method? Since the full functionality of different parts of the human brain is still not known completely, grouping based on the AAL map seems a logical choice, considering that it is regularly used in fMRI studies. Our future work will aim at evaluating the correlation of the EEG-based activity localization with that of fMRI.

**ACKNOWLEDGMENT**

This project was supported by the Computational and Simulation Sciences Transformational Capability Platform of Commonwealth Scientific and Industrial Research Organisation (CSIRO), Australia, along with University of New South Wales (UNSW), Canberra, Australia. The authors would like to thank Dr. Chao Suo and Dr. Roger Koenig-Robert of Monash Biomedical Imaging (MBI) Laboratory, VIC, Australia for their valuable comments and suggestions.